\documentclass[pre,aps,twocolumn,showpacs,amsmath,amssymb,amsfonts]{revtex4}
\usepackage{epsfig}
\usepackage{amscd}
\begin{document}
\title{Nontrivial critical crossover between directed percolation models:
Effect of infinitely many absorbing states}
\author{Su-Chan Park}
\affiliation{Institut f\"ur Theoretische Physik, Universit\"at zu K\"oln,
Z\"ulpicher Strasse 77, 50937 K\"oln, Germany}
\author{Hyunggyu Park}
\affiliation{School of Physics, Korea Institute for Advanced Study, Seoul 130-722, Korea}
\date{\today}
\begin{abstract}
At non-equilibrium phase transitions into absorbing (trapped)
states, it is well known that the directed percolation (DP)
critical scaling is shared by two classes of models with a single
(S) absorbing state and with infinitely many (IM) absorbing
states. We study the crossover behavior in one dimension, arising
from a considerable reduction of the number of absorbing states
(typically from the IM-type to the S-type DP models), by following
two different (excitatory or inhibitory) routes which make the
auxiliary field density abruptly jump at the crossover. Along the
excitatory route, the system becomes overly activated even for an
infinitesimal perturbation and its crossover becomes
discontinuous. Along the inhibitory route, we find continuous
crossover with the universal crossover exponent $\phi\simeq
1.78(6)$, which is argued to be equal to $\nu_\|$, the relaxation
time exponent of the DP universality class on a general footing.
This conjecture is also confirmed in the case of the directed
Ising (parity-conserving) class. Finally, we discuss the effect of
diffusion to the IM-type models and suggest an argument why
diffusive models with some hybrid-type reactions should belong to
the DP class.
\end{abstract}
\maketitle
\section{\label{Sec:intro}introduction}

The directed percolation (DP) has been studied extensively as one
of typical dynamic critical phenomenon far from equilibrium
\cite{H00}. Nonequilibrium phase transitions of systems with a
unique absorbing state are found to belong to the DP class if no
symmetry or conservation of the order parameter is present
\cite{DPC}. Even systems with infinitely many (IM) absorbing
states such as the pair contact process (PCP) \cite{J93} are
believed to share the same critical behavior with the DP, but the
theoretical understanding of the IM-type DP (DP${}_\textrm{IM}$) models  is
still lacking. For example, the phenomenological field theory
introduced in Ref.~\cite{FT_PCP} and elaborated in Ref.~\cite{W02}
shows inconsistency with the numerical studies. In fact, the field
theory based on the phenomenological Langevin equation predicts
that the PCP should belong to the dynamic percolation rather than
the DP class \cite{W02}. Besides the universality issue, its
spreading dynamics is also not fully understood \cite{GCR,DH}.

One may reduce the number of absorbing states significantly by
introducing particle diffusion (PCP with diffusion or PCPD)
\cite{HH04}. Surprisingly, the PCPD has brought up serious
turmoil, which could not be settled down in spite of extensive
numerical \cite{KC03,NP04,H06} and analytical \cite{JvWOT04}
studies. The answer seems to be one of two possibilities: The PCPD
belongs to the DP class with a long transient, or forms a new
universality class distinct from known universality classes to
date. As an attempt to resolve the issue, the present authors
suggested two different approaches to the PCPD.

First, introducing the dynamic perturbation which is implemented
by the biased hopping, we showed that the one-dimensional PCPD
with biased diffusion (driven PCPD or DPCPD), exhibits critical
scaling distinct from the unbiased PCPD, but instead shares the
critical behavior with the two-dimensional PCPD without bias
(dimensional reduction) \cite{PP05a,PP05b}. Since all known DP
models are robust against the biased diffusion \cite{PP05a}, a
DPCPD-type variant can serve as a litmus test for the PCPD scaling
\cite{P06}. Second, we  studied the crossover behavior from the
PCPD to the DP by introducing single-particle
annihilation/branching reactions and showed that there are
diverging crossover scales with the universal nontrivial crossover
exponent \cite{PP06}. These results provided another evidence that
the PCPD is distinct from the DP.

In this paper, we study the crossover behavior from the DP${}_\textrm{IM}$ to
the DP in order to understand better the difference between these
two ``equivalent'' (DP) universality classes. Actually, it would
be absurd to talk about the crossover between two models belonging
to the identical universality class (see
Sec.~\ref{Sec:single_to_single}). However, the DP${}_\textrm{IM}$ models differ
from the DP models in regard to the ``non-order'' parameter at the
transitions, which shows a discontinuous singularity at the
crossover. This singularity induces a well-defined and nontrivial
crossover from the DP${}_\textrm{IM}$ to the DP, more generally a crossover
arising from a considerable reduction of the number of absorbing
states (between two different DP${}_\textrm{IM}$).

We find two distinct crossover behaviors depending on the routes
to reduce the number of absorbing states. Infinitesimal inclusion
of an ``excitatory'' process (like single-particle branching)
makes the system overly active, which gives rise to a
discontinuous crossover (see Sec.~\ref{Sec:dpima}). While, the
opposite ``inhibitory'' route (like single-particle annihilation)
reveals a continuous crossover with the nontrivial crossover
exponent $\phi\simeq 1.78(6)$. We find that this crossover
exponent is universal for various kinds of models including the
PCP and the triplet contact process (TCP). We argue that the
crossover exponent is not independent but equal to $\nu_\|$, the
relaxation time exponent of the DP universality class on a general
footing. This conjecture is also confirmed in the case of the
directed Ising (parity-conserving) class (see Sec.~\ref{Sec:discussion}).

The PCPD {\it per se} can be considered one of crossover models
from the DP${}_\textrm{IM}$ by allowing diffusion, which reduces the number of
absorbing states considerably (from exponentially many to linearly
many absorbing states with respect to system size). However, the
crossover study from the PCP to the PCPD does not give any useful
information on the DP${}_\textrm{IM}$, because the particle diffusion makes the
system more active (excitatory) and the crossover  turns out to be
discontinuous.

Finally, we study the crossover from the diffusive reaction models
to the DP. Such an example is the crossover from the PCPD to the
DP studied in Ref.~\cite{PP06}. In Sec.~\ref{Sec:dptodp}, the
hybrid-type models with diffusion ({\tt tp12}:
$2A\rightarrow\emptyset, 3A\rightarrow 4A$) \cite{KC03} are
perturbed by adding a single-particle annihilation process
($A\rightarrow\emptyset$) and its crossover to the DP is
investigated. These hybrid-type models (where the branching
process is of higher order than the annihilation process) are
numerically known to belong to the DP class. As expected, the
critical line emanates ``linearly'' from the hybrid-type model
point. We suggest an argument why these models should belong to
the DP rather than a PCPD-type nontrivial class.

\section{\label{Sec:single_to_single} Crossover between the identical universality class?}
This section considers a $d$-dimensional stochastic system
of hard core particles with dynamics summarized in Table~\ref{Table:DP}.
In Ref.~\cite{PP05c}, it is shown that two
different stochastic systems modeled by tilded and untilded rates
are equivalent if transition rates satisfy the relations
\begin{equation}
\begin{aligned}
\tilde \sigma = \mu \sigma&,\qquad 2  \tilde D + \tilde \xi = 2  D + \xi,\\
\tilde D + \frac{\tilde \sigma}{2} &= D + \frac{\sigma}{2},\\
\tilde \lambda + \frac{\tilde \eta}{2} + \frac{\tilde \sigma}{2} &=
\frac{1}{\mu}\left (\lambda + \frac{\eta}{2}+\frac{\sigma}{2}\right ),\\
\tilde \lambda + \tilde \eta + \tilde \sigma - 2 \tilde D  &= \lambda +
\eta+\sigma - 2 D ,
\end{aligned}
\label{Eq:prime_unprime}
\end{equation}
where any positive number $\mu$ which renders all tilded and
untilded transition rates be nonnegative is physically meaningful.
By equivalence is meant that all correlation functions of the
model with tilded parameters can be deduced from those of the
model with untilded parameters and vice versa. For instance, the
particle density $\rho$ at time $t$ of two different stochastic
many body systems becomes \cite{PP05c}
\begin{equation}
\rho(D,\lambda,\eta,\xi,\sigma;\rho_0,t) = \frac{1}{\mu} \tilde \rho(\tilde D,
\tilde \lambda,\tilde \eta,\tilde \xi,\tilde \sigma;\tilde \rho_0,t),
\label{Eq:rho_relation}
\end{equation}
if the initial density of two systems has the relation $\tilde
\rho_0 = \mu \rho_0$.  Needless to say, both $\rho_0$ and $\tilde
\rho_0$ should lie between 0 and 1. Although the equivalence is
shown only for the one-dimensional systems in Ref.~\cite{PP05c},
Eq.~\eqref{Eq:prime_unprime} is generally true for any
$d$-dimensional systems, which can be easily shown by the same
technique developed in Ref.~\cite{PP05c}.

\begin{table}[t]
\caption{\label{Table:DP} $d$-dimensional reaction-diffusion
processes of single species with hard core exclusion and their
rates.}
\begin{ruledtabular}
\begin{tabular}{ccl}
\text{Diffusion}& $A\emptyset \leftrightarrow \emptyset A$ &
\text{with
rate }$D/d$\\
\text{Pair annihilation}&$ A A \rightarrow \emptyset \emptyset$  &
\text{with rate }$\lambda/d$\\
\text{Coalescence} & $ A A \rightarrow A \emptyset$  &
\text{with rate }$ {\eta}/(2d)$ \\
\text{Coalescence} & $ A A \rightarrow \emptyset A$
&\text{with rate } $ {\eta}/(2d) $\\
\text{Death} &  $A   \rightarrow \emptyset$
&\text{with rate } $\xi$ \\
\text{Branching}  &  $\emptyset A \rightarrow A A $
&\text{with rate } ${\sigma}/(2d)$ \\
\text{Branching}  & $ A \emptyset  \rightarrow A A $&
\text{with rate } $\sigma/(2d) $\\
\end{tabular}
\end{ruledtabular}
\end{table}
Now consider the branching annihilating random walks with one
offspring (BAW1) \cite{TT92} which corresponds to the model with
$\xi = 0$ in Table~\ref{Table:DP}. The parameters used in Ref.~\cite{TT92}
in one dimension are $D = p/2$, $\lambda = p$, and
$\eta = \sigma = 1-p$ with the tuning parameter $p$.  If $w \equiv
\mu - 1$ is very small and nonnegative, the solution of
Eq.~\eqref{Eq:prime_unprime} up to the order of $w$ is
\begin{equation}
\begin{aligned}
&\tilde D -D=  - \frac{w}{2} \sigma,\quad \tilde \sigma -\sigma =
w \sigma,\quad \tilde \xi = w \sigma,\\
&\tilde \lambda - \lambda  \simeq -  w (\eta+ 2 \lambda),\quad
\tilde \eta - \eta \simeq  w (\eta + 2 \lambda - 2 \sigma).
\end{aligned}
\label{Eq:crossBAW}
\end{equation}
If $w$ is sufficiently small, it is always possible to associate
the BAW1 with a stochastic process with spontaneous death in an
equivalent way with all nonnegative rates. Since the BAW1 in high
dimensions is also known to have a nontrivial transition point
\cite{CCD04}, the following discussion is valid in any spatial
dimension. The transition points for stochastic systems with small
$w$ can be always calculated exactly from
Eq.~\eqref{Eq:prime_unprime} (approximately from
Eq.~\eqref{Eq:crossBAW}), if the transition point of the BAW1 is
given.

The conclusions from the above analysis are two-fold. First, it is
clear from Eq.~\eqref{Eq:crossBAW} that the phase boundary
(critical line) should meet the BAW1 transition point {\em
linearly} with finite slope, as $w$ vanishes. This implies that
there is no additional singularity involved near $w=0$, which is
fully expected from the crossover between models with the
identical universality class. If one defines the crossover
exponent $\phi$ from the shape of the critical line near $w=0$
(see Sec.~\ref{Sec:dpima}), one can say that $\phi=1$. Second, the
critical decay of the density is given by $\tilde \rho_c(t) =
(1+w) \rho_c(t)$ from Eq.~\eqref{Eq:rho_relation}, which implies
that there is no diverging crossover time scale for small $w$.

Since the introduction of the spontaneous death does not change
the structure of the absorbing phase space (single absorbing
state) let alone the universality class, the above analysis is in
good harmony with the naive expectation as to the ``crossover''
between two models belonging to the identical class. In the next
section, however, we will show that the substantial change of the
absorbing phase space without affecting the universality class
will trigger a nontrivial crossover.
\section{\label{Sec:dpima}Crossover from the DP${}_\textrm{IM}$ to the DP}
Unlike the BAW1, the pair contact process (PCP) is the prototype
of the DP${}_\textrm{IM}$ models with exponentially many absorbing states. By
introducing single-particle reactions to the PCP, the number of
absorbing states changes drastically from infinity to one. This
section shows that this qualitative change is reflected into the
singular behavior of the phase boundary close to the PCP
transition point in one dimension.

The dynamics of the model is summarized as
\begin{subequations}
\label{Eq:cross_PCP}
\begin{eqnarray}
\label{Eq:cross_PCP_a}
&&AA \stackrel{p}{\longrightarrow} \emptyset \emptyset,\quad
\left .\begin{matrix} AA\emptyset\\ \emptyset AA \end{matrix} \right \}
\begin{CD}
@>(1-p)/2>> AAA,
\end{CD} \\
&&\begin{CD}
\left . \begin{matrix} A\emptyset\\ \emptyset A \end{matrix} \right \}
@>(1-q) w/2>> \emptyset \emptyset,\quad
\left . \begin{matrix} A\emptyset\\ \emptyset A \end{matrix} \right \}
@>q w/2>> AA,
\end{CD}
\label{Eq:cross_PCP_b}
\end{eqnarray}
\end{subequations}
\noindent
where $0\le q \le 1$. For the PCP at $w=0$, any configuration
without a pair of neighboring particles (a mixture of isolated
particles and vacant sites) is absorbing and its number grows
exponentially with system size. The order parameter of the PCP is
the pair density (the number density of $AA$ pairs) and the
particle density field is {\em auxiliary} which is finite even in
the absorbing phase. At nonzero $w$, an isolated particle becomes
active and only the vacuum becomes the true absorbing state. In
this case, the particle density is usually adopted as the order
parameter and the pair density scales in the same way.

\begin{figure}[t]
\includegraphics[width=0.45\textwidth]{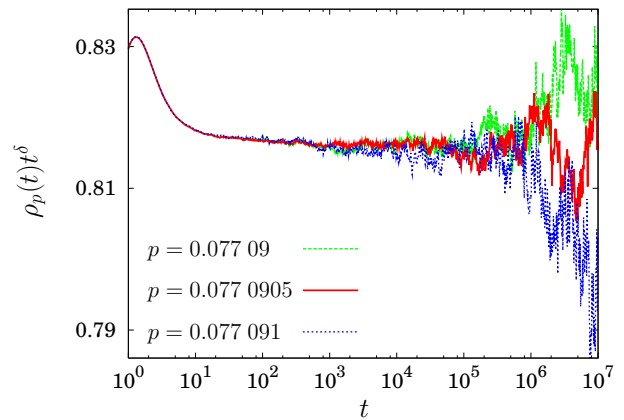}
\caption{\label{Fig:PCP_pc} (Color online) Semilogarithmic plot of
$\rho_p(t) t^{\delta}$ vs $t$ of the PCP near criticality with
$\delta = 0.1595$ to be the exponent of the DP. Since the upper
(lower) curve veers up (down), we estimate the critical point
as $p_0 = 0.077\;0905(5)$ with the error in the last digit by 5.}
\end{figure}

Figure~\ref{Fig:PCP_pc} locates the transition point of the PCP
($w=0$) at $p_0 = 0.077\;0905(5)$ by exploiting the critical decay
of the pair density  as $\rho_p(t) \sim t^{-\delta}$ with $\delta$
to be the critical exponent of the DP class whose accurate value
can be found in \cite{J99}.   In numerical simulations, the system
size is $L=2^{18}$ and the number of independent samples are 750,
1500, and 400 for the data in the active, critical, and absorbing
phases, respectively. The flatness of $\rho_p(t) t^{\delta}$ over
four log decades in time confirms the solid DP critical scaling of
the PCP.

At finite $w$, the model still belongs to the DP class
irrespective of $q$. Unlike the PCPD to the DP crossover model in
Ref.~\cite{PP06}, however, the critical lines show two completely
different singular behaviors, depending on the value of $q$. For
large $q$, the activity of the system is enhanced by additional
single-particle reaction processes (excitatory process) and the
system becomes overly activated even with infinitesimal $w$. The
critical line does not converge to the PCP critical point as $w$
decreases to zero ($w=0^+$) and shows a discontinuous jump. On the
other hand, for small $q$, the system activity is suppressed
(inhibitory process) and the system becomes more inactive. The
critical line nicely converges to the PCP critical point and shows
a continuous crossover with a nontrivial crossover exponent.

\begin{figure}[b]
\includegraphics[width=0.48\textwidth]{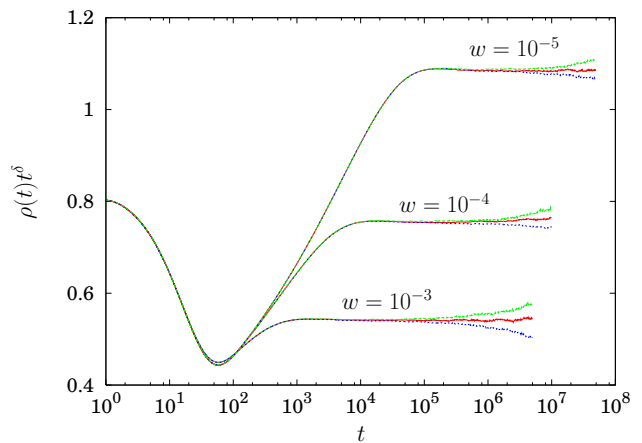}
\caption{\label{Fig:A2A} (Color online) Plots of $\rho(t)
t^\delta$ vs $t$ for $w=10^{-3}$, $10^{-4}$, and $w=10^{-5}$ at
$q=1$ close to the critical points in semi-logarithmic scales.
Again, $\delta$ assumes the DP value. The curves in the middle
correspond to $p=0.1451$, $0.1448$, and $0.1448$ (from bottom to
top), respectively and the value of $p$'s of other two curves are
$\pm 0.0001$ off from the middle value. As $w$ becomes smaller,
the relaxation time becomes larger though the critical point does
not change much and approaches $p \simeq 0.1448$. }
\end{figure}
First, we choose the $q=1$ case as a typical excitatory route of
the crossover from the DP${}_\textrm{IM}$ to the DP. As shown in
Fig.~\ref{Fig:A2A}, the critical line approaches $p\simeq  0.1448$
as $w$ approaches zero, which is by far above the critical value
of the PCP ($p_0\simeq0.077$).  So there is a big jump of the
critical line at $w=0$. The discontinuity can be understood as
follows: Consider a system with $p$ slightly above the PCP
critical point $p_0$ and $0<w\ll\tau^{-1}$ where $\tau$ is the
relaxation time which is finite off criticality.
Then the single-particle branching event ($A\rightarrow 2A$) with
the characteristic time of $w^{-1}$  occurs effectively after the
system falls into one of the PCP absorbing states in which the
isolated particle density is finite. Since the branching event
creates a new pair, the system is reactivated and performs the
damage-spreading-type ``defect dynamics'' for some time
proportional to $\tau$ and again falls into one of the PCP
absorbing states. This defect dynamics continues forever with the
period of time $w^{-1}$. As the particle density is finite (and
quite large) even in the PCP absorbing states, the time-averaged
particle density in this iterated process should be finite in this
region of the phase diagram. This implies that the continuous
absorbing phase transition at infinitesimal $w$ into vacuum should
occur way above $p_0$, which is consistent with our finding. Note
that the discontinuity in the auxiliary field (particle) density
is crucial in this crossover.

Actually, the same argument can be applied to the crossover from
the PCP to the PCPD. We can introduce the diffusion  rather than
the single-particle reactions and again consider $p$ slightly
above $p_0$.  Let $\rho_0$ denote the isolated particle density at
the PCP absorbing states, then the characteristic length scale
between isolated particles is $1/\rho_0$. If $0< D\rho_0^2 \ll
\tau^{-1}$ with the diffusion constant $D$, the ``defect
dynamics'' will continue again indefinitely for small $D$. So the
phase boundary in the $D-p$ plane should have a discontinuity at
$D=0$.

\begin{table}[b]
\caption{\label{Table:PCP_pc} Critical points of the model with
dynamics of Eq.~\eqref{Eq:cross_PCP} for some values of $w$'s at
$q=0$. The numbers in the parentheses indicate the error of the
last digits.}
\begin{ruledtabular}
\begin{tabular}{rl}
$w$&$p_c(w)$\\
\hline
$0              $&$0.077~0905 (5)$\\
$10^{-5}        $&$0.077~002  (3)$\\
$5\times 10^{-5}$&$0.076~885  (3)$\\
$10^{-4}        $&$0.076~784  (4)$\\
$2\times 10^{-4}$&$0.076~642  (2)$\\
$3\times 10^{-4}$&$0.076~530  (5)$\\
$4\times 10^{-4}$&$0.076~432  (2)$\\
$5\times 10^{-4}$&$0.076~345  (2)$\\
$6\times 10^{-4}$&$0.076~264  (1)$\\
$10^{-3}        $&$0.075~988  (4)$
\end{tabular}
\end{ruledtabular}
\end{table}

\begin{figure}[t]
\includegraphics[width=0.45\textwidth]{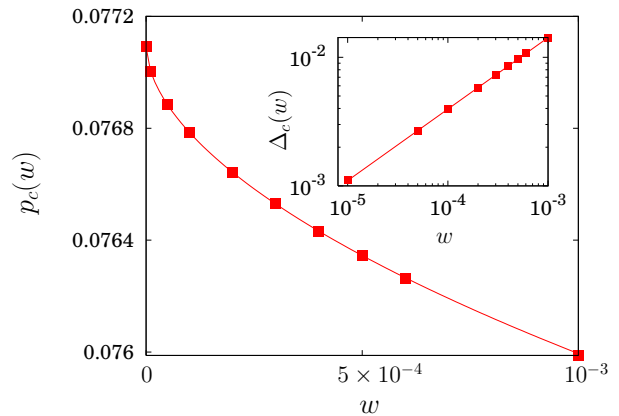}
\caption{\label{Fig:PCPphase} (Color online) Phase boundary of the
model of Eq.~\eqref{Eq:cross_PCP} at $q=0$ in $w-p$ plane. Symbols
locate the numerically estimated critical points. The error of the
critical point is smaller than the symbol size. The curve shows
the least-square-fit result of the phase boundary. The absorbing
(active) phase is above (below) the curve. Inset: the same but the
vertical axis is $\Delta_c(w)$ in log-log scale. The slope
corresponds to the inverse of the crossover exponent which is
estimated as $\phi = 1.78(6)$. }
\end{figure}

Let us turn to the crossover model with $q=0$, which should
represent a typical inhibitory route.  Table~\ref{Table:PCP_pc}
summarizes the critical points of the model for some $w$'s at
$q=0$ and the corresponding phase boundary is plotted in
Fig.~\ref{Fig:PCPphase}. Unlike the previous case, the reactive
phase shrinks continuously with the rate of additional
single-particle annihilation process and the phase boundary is
continuous. The usual analysis method can be applied to this case
\cite{DL9}. If we define $\Delta = (p_0 - p)/p_0$ and $\Delta_c(w)
= (p_0 - p_c(w))/p_0$, the phase boundary is well fitted by
$\Delta_c \sim w^{1/\phi}$ with $\phi^{-1} = 0.56(2)$ or $\phi =
1.78(6)$; see the inset of Fig.~\ref{Fig:PCPphase}. Let us assume
the existence of the  well-defined crossover scaling which is
described by the scaling function \cite{DL9}
\begin{equation}
\rho_p(\Delta,w;t) = t^{-\delta} {\cal F}(\Delta^{\nu_\|} t,
w^{\mu_\|} t),
\label{Eq:scaling}
\end{equation}
where $\rho_p$ is the pair density and $\mu_\| = \nu_\|/\phi$ with
$\phi$ estimated in the above. We examine whether the scaling
function in Eq.~\eqref{Eq:scaling} correctly describe the
crossover near the PCP critical point.

\begin{figure}[b]
\includegraphics[width=0.45\textwidth]{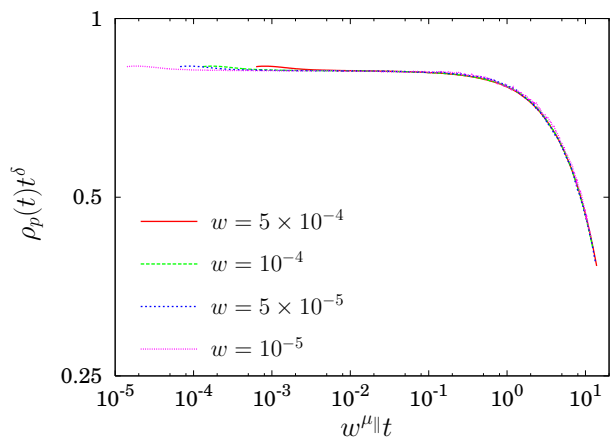}
\caption{\label{Fig:mucollapse} (Color online) Log-log plot of the
scaling function Eq.~\eqref{Eq:mucoll} for the PCP crossover model
using $\delta = 0.1595$ and $\mu = 0.97$. All curves are collapsed
into a single curve. }
\end{figure}
\begin{figure}[t]
\includegraphics[width=0.45\textwidth]{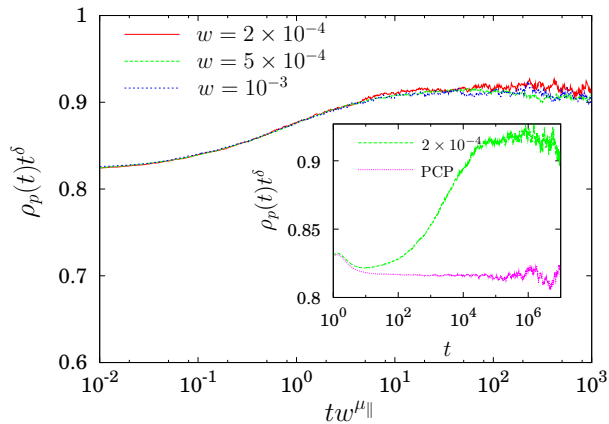}
\caption{\label{Fig:pair} (Color online) Scaling plot of
$\rho_p(t) t^{\delta}$ vs $w^{\mu_\|}t$ with $\delta = 0.1595$ and
$\mu_\|=0.97$ at criticality in semi-log scales. Inset: Scaling
function ${\cal H}$ at finite $w$ and $w=0$. The asymptotic value
of $w=0$ is  different from the $w\rightarrow 0$ limit. }
\end{figure}

First, we measure the pair density for various $w$'s at
$\Delta=0$. From the scaling ansatz \eqref{Eq:scaling}, the pair
density at $\Delta = 0$ should collapse as
\begin{equation}
t^{\delta} \rho_p(t) = {\cal G}(w^{\mu_\|} t ).
\label{Eq:mucoll}
\end{equation}
With $\mu_\| \simeq 0.97$, all curves for the pair density are
collapsed into a single curve as Fig.~\ref{Fig:mucollapse} shows.
Next, we take $\Delta = \Delta_c(w)$ along the critical line.
Since $\Delta_c(w) \simeq w^{1/\phi}$ and $\nu_\|/\phi = \mu_\|$,
the scaling function should take the form
\begin{equation}
t^{\delta} \rho_p(\Delta_c(w);t) = {\cal H}(w^{\mu_\|} t ),
\label{Eq:DCcoll}
\end{equation}
where ${\cal H}(x)$ approaches a constant as $x \rightarrow
\infty$. In Fig.~\ref{Fig:pair}, all curves at different critical
points collapse well into a single curve. Hence we conclude that
the scaling function Eq.~\eqref{Eq:scaling} correctly describes
the crossover behavior from the DP${}_\textrm{IM}$ to the the DP.

\begin{figure}[b]
\includegraphics[width=0.45\textwidth]{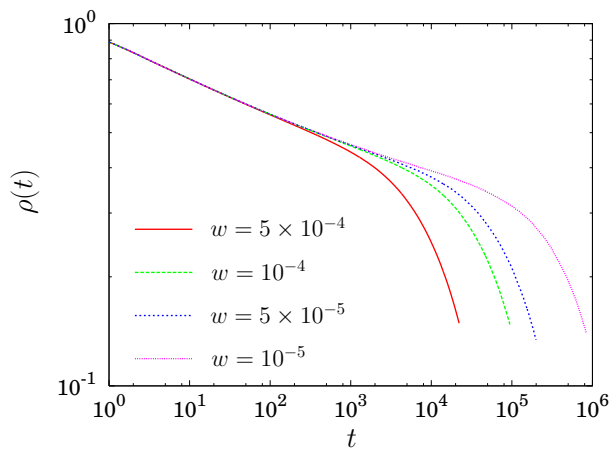}
\caption{\label{Fig:single} (Color online) Log-log plot of the
particle density for some finite $w$'s at $\Delta = 0$. For any
finite value of $w$, the particle density $\rho(t)$ will go to
zero, which is clearly different from the $w=0$ case. }
\end{figure}
Since the models at  $w=0$ and at $w\neq 0$ belong to the same DP
universality class, it is natural to ask what is the origin of
such a nontrivial singularity near the PCP critical point. The
inset of Fig.~\ref{Fig:pair} gives a hint to this question, which
shows that the scaling function ${\cal H}$ (the amplitude of the
critical decay) does not approach the PCP value as $w$ goes to
zero, i.e., it is not continuous at $w=0$. The discontinuity in
this amplitude must originate again from the discontinuity in the
auxiliary field (particle) density.  One can see it directly from
the behavior of the particle density ($\rho$). Unlike the pair
density, $\rho$ can not be described by the scaling function
\eqref{Eq:scaling}. Consider again the case at $\Delta = 0$ and
nonzero $w$. For any finite value of $w$ in the thermodynamic
limit, $\rho(t)$ approaches to zero as $t\rightarrow \infty$; see
Fig.~\ref{Fig:single}. On the other hand, the model at $\Delta =
w=0$ (the critical PCP), has a nonzero density of $\rho(t)$ as
$t\rightarrow \infty$. In other words, the $w\rightarrow 0$
limiting process is different from the $w=0$ model itself in
regard to the auxiliary field density.

To check the universality of  the crossover exponent, we study the
modified PCP with the replacement of $2A \rightarrow 3A$ with $3A
\rightarrow 4A$ in Eq.~\eqref{Eq:cross_PCP} which is the model of
Eq.~\eqref{Eq:tp12} with no diffusion ($D=0$). This model  also
has infinitely many absorbing states and belongs to the DP${}_\textrm{IM}$
class. By introducing single-particle reactions ($w\neq 0$), the
same crossover behavior is found as the above (data not shown).

\begin{figure}[t]
\includegraphics[width=0.45\textwidth]{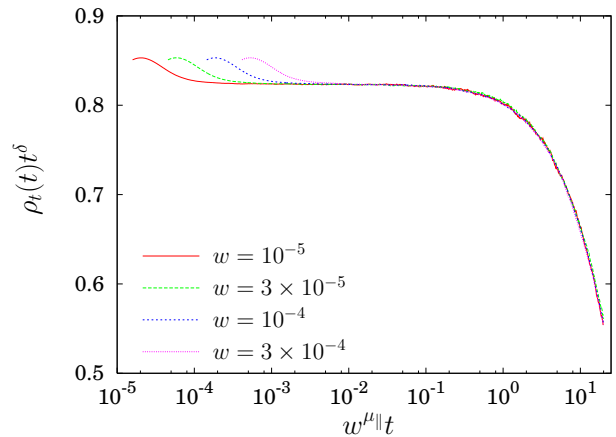}
\caption{\label{Fig:TCPcollapse} (Color online) Scaling plot of
$\rho_t(t) t^{\delta}$ vs $w^{\mu_\|} t$ with the same exponents
in Fig.~\ref{Fig:mucollapse} for the crossover TCP model of
Eq.~\eqref{Eq:TCPcross} in semi-log scales. As in
Fig.~\ref{Fig:mucollapse}, all curves are collapsed into a single
curve.}
\end{figure}

We also study more general crossover behavior from one DP${}_\textrm{IM}$ to
another DP${}_\textrm{IM}$ with the considerably reduced number of absorbing
states. To be specific, we consider the triplet contact process
(TCP) and its crossover model by introducing the $2A \rightarrow
A$ process without spontaneous death. The TCP with pair dynamics
is defined as
\begin{equation}
\label{Eq:TCPcross}
\begin{aligned}
&AAA \stackrel{p}{\longrightarrow} \emptyset \emptyset\emptyset,\quad
\left .\begin{matrix} AAA\emptyset\\ \emptyset AAA \end{matrix} \right \}
\begin{CD}
@>(1-p)/2>> AAAA,
\end{CD}\\
&\begin{CD}
 AA\emptyset
@>w/2>> A\emptyset \emptyset,\quad
\emptyset AA @>w/2>> \emptyset \emptyset A.
\end{CD}
\end{aligned}
\end{equation}
The above model has infinitely many absorbing states, but with
nonzero $w$ the number of absorbing states is greatly reduced. At
$w=0$, there is again a jump in the auxiliary field density (here,
the pair density). We found that the critical point for the TCP at
$w=0$ is $p_c = 0.036~865(5)$, exploiting the DP critical scaling
(data not shown). Figure~\ref{Fig:TCPcollapse} shows the scaling
plot of the triplet density $\rho_t$ in the same way as in
Fig.~\ref{Fig:mucollapse}. We also measured the crossover exponent
from the phase boundary and found the same exponent (data not
shown).

Hence we conclude that there is the well-defined and universal
crossover scaling from the DP${}_\textrm{IM}$ to the DP which is mediated by
the significant reduction of the number of absorbing states. The
discontinuity in the auxiliary field plays a crucial role in this
nontrivial crossover.

\section{\label{Sec:discussion} Conjecture on the crossover exponent $\phi$}
The crossover exponent from the DP${}_\textrm{IM}$ to the DP is
estimated as $\phi = 1.78(6)$. Since this crossover occurs between
the same universality class, we are suspicious that $\phi$ may not
be independent but  related to the well-known DP critical
exponents. Actually, we argue that the crossover exponent is given
by the DP relaxation time exponent: $\phi = \nu_\|\simeq 1.733$
(or $\mu_\|=1$) which is compatible with the numerical estimation
within error. The reason is as follows: Take the model of
Eq.~\eqref{Eq:cross_PCP} at $q=0$. The critical line should be
determined by the competition between the single-particle
annihilation process ($A\rightarrow\emptyset$) parameterized by
$w$ and the multi-particle (pair) reaction process
($2A\rightarrow\emptyset$ or $3A$) parameterized by
$\Delta\sim(p_0-p)$. We expect both events should appear at the
same time scale along the critical line to balance off each other.
Since the single-particle (auxiliary field) density is finite at
the PCP (DP${}_\textrm{IM}$) point, the time scale for
$A\rightarrow\emptyset$ should be simply proportional to $w^{-1}$.
The time scale for the pair reaction process should be given by
the relaxation time scale $\tau\sim \Delta^{-\nu_\|}$.
Consequently, the critical line is determined as
$\Delta_c\sim(p_0-p_c(w)) \sim w^{1/\nu_\|}$, which yields
$\phi=\nu_\|$ and equivalently $\mu_\|=1$.

Considering the crossover between the DP models with a unique
absorbing state, the time scale for the process parameterized by
$w$ is proportional to $w^{-\nu_\|}$ like the other competing
process because the auxiliary field (particle) density is also
vanishing critically as $w$ decreases to zero. In this case, we
get $\phi=1$ and equivalently $\mu_\|=\nu_\|$, which is consistent
with our result in Sec.~\ref{Sec:single_to_single}.

\begin{table}[b]
\caption{\label{Table:DI_pc} Critical points of the DI crossover
model for some values of $w$'s (see the text). The numbers in the
parentheses indicate the error of the last digits.}
\begin{ruledtabular}
\begin{tabular}{rl}
$w$&$p_c(w)$\\
\hline
$0              $&$0.4518     (1)$\\
$10^{-4}        $&$0.4443     (2)$\\
$3\times 10^{-4}$&$0.4405     (2)$\\
$10^{-3}        $&$0.4353     (1)$\\
$3\times 10^{-3}$&$0.4276     (1)$\\
\end{tabular}
\end{ruledtabular}
\end{table}
Since our  argument for the crossover exponent is generally
applicable to any universality class, we can check its validity
through studying the similar type crossover between the directed
Ising (DI) class models \cite{DIPark}. Consider a one-dimensional
system with two species, say $A$ and $B$. Between the same
species, hard core exclusion is applied, but different species can
reside at the same site. The dynamic rules are as follows: The
dynamics always starts with an $A$ particle. A randomly chosen $A$
particle can hop to one of nearest neighbors with probability $p$.
If two $A$ particles meet at the same site by hopping, both
particles are removed with probability $\frac{1}{2}$. If this
annihilation attempt fails, the particle goes back to the original
site. With probability $1-p$, a $B$ particle is generated at the
same site occupied by the chosen $A$ particle. If that site is
already occupied by another $B$ particle, the two $B$ particles
transmute to two $A$ particles which will be placed at two nearest
neighbor sites. If any of transmuted $A$ particles is placed at
the site already occupied by another $A$ particle, both particles
are annihilated immediately. In summary, $2A\rightarrow\emptyset$,
$A\rightarrow A+B$, and $2B\rightarrow 2A$ processes are allowed
with $A$ particle diffusion.

Since $B$ particles are not allowed to hop, the system is inactive
without an $A$ particle but only with $B$ particles. The number of
the absorbing states grows exponentially with system size and the
auxiliary field ($B$ particle) density is finite at the absorbing
transition. Besides, the number of $A$ particles is conserved
modulo 2, which is the characteristic of the DI (or
parity-conserving) class.

As in Sec.~\ref{Sec:dpima}, we study the crossover by introducing
spontaneous annihilation of $B$ particles (inhibitory route) with
rate $w$. As expected, we find the DI critical scaling for both
$w=0$ and $w\neq 0$ cases. We locate the critical line by
exploiting the known DI critical exponents \cite{DIPark}, which is
summarized in Table~\ref{Table:DI_pc} for some $w$'s. Since the
critical exponent $\nu_\|$ of the DI class ($\simeq 3.25$) is much
larger than that of the DP ($\simeq 1.733$), the accuracy of the
critical points in Table~\ref{Table:DI_pc} is worse than that for
the DP cases. From these data, one can estimate the crossover
exponent $1/\phi_\text{DI}$ as $0.34(4)$ which should be compared
with $1/\nu_\|$ of the DI class ($\simeq 0.31$). Hence we conclude
that our argument for $\phi=\nu_\|$ also applies to the crossover
from the IM-type DI to the DI models.

\section{\label{Sec:dptodp}Diffusion effect}
This section studies how the crossover scaling is affected if
particles are allowed to hop in models considered in
Sec.~\ref{Sec:dpima}. With single-particle diffusion, the PCP
becomes the PCPD where the particle (auxiliary field) density as
well as the pair density vanishes at criticality even without any
single-particle reaction process. The crossover from the PCPD to
the DP caused by including single-particle reactions has been
studied previously by the present authors \cite{PP06} where the
value of the crossover exponent is reported as $1/\phi = 0.58(3)$.
This value is quite close to that obtained for the crossover from
the DP${}_\textrm{IM}$ to the DP in Sec.~\ref{Sec:dpima}. This
similarity may mislead one to jump to the wrong conclusion that
the critical nature of the PCP and the PCPD is equivalent.
However, one should remember that the origin of the nontrivial
crossover from the PCP to the DP  lies in the finiteness of the
auxiliary field density at the PCP critical point, while it
vanishes at the PCPD critical point. So, if the PCPD belongs to
the DP class, then one should expect a trivial crossover with
$\phi=1$. Our finding of the nontrivial crossover in \cite{PP06}
implies that the PCPD class is distinct from the DP class. Hence
it is likely that the similarity of two crossover exponents be a
mere coincidence.

The distinction of the PCPD from the DP can also be evidenced by
the study of the crossover model with the hybrid-type reaction
dynamics $2A \rightarrow 0$ and $3A \rightarrow 4A$. Without
diffusion, this model belongs to the DP${}_\textrm{IM}$  and its crossover to
the DP was studied in Sec.~\ref{Sec:dpima}. Even if particles are
allowed to diffuse, this model (called  as {\tt tp12}) is numerically
known to belong to the DP class \cite{KC03}.

\begin{figure}[t]
\includegraphics[width=0.45\textwidth]{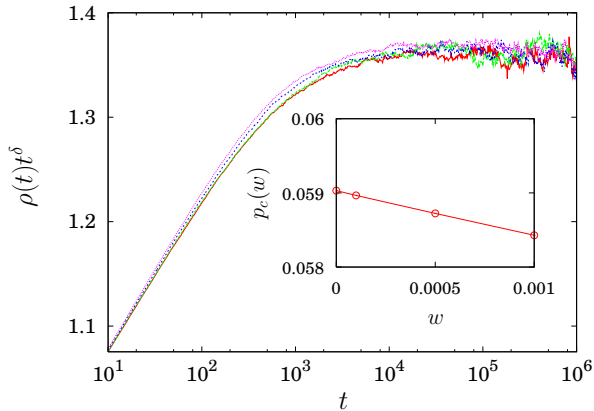}
\caption{\label{Fig:tp12} (Color online) Semi-log plot of $\rho(t)
t^{\delta}$ vs $t$ with $\delta = 0.1595$ at $p_c
=0.058~427,~0.058~723,~0.058~966$, and $0.059~031$ for
$w=10^{-3},~5\times 10^{-4},~10^{-4}$, and $0$ (from left to
right), respectively. There is no diverging time scale as $w$
approaches zero. Inset: Phase boundary near $w=0$. Each symbol
corresponds to the critical point used in the main figure. The
straight line is drawn between two consecutive points,
independently. }
\end{figure}
It will be illuminating to see how the diffusion in the {\tt tp12} can
change the crossover behavior to the DP. The dynamics of the model
is summarized as
\begin{equation}
\label{Eq:tp12}
\begin{aligned}
&AA \stackrel{p}{\longrightarrow} \emptyset \emptyset,\quad
\left .\begin{matrix} AAA\emptyset\\ \emptyset AAA \end{matrix} \right \}
\begin{CD}
@>(1-p)/2>> AAAA,
\end{CD} \\
&\begin{CD}
\left . \begin{matrix} A\emptyset\\ \emptyset A \end{matrix} \right \}
@>w/2>> \emptyset \emptyset,\quad
A\emptyset \stackrel{D}{\longleftrightarrow} \emptyset A,
\end{CD}
\end{aligned}
\end{equation}
where $D = (1-w)/2$. The model at $w=0$ is the {\tt tp12}. Our
numerical results in Fig.~\ref{Fig:tp12} show the typical
``crossover'' behavior between the identical universality class
discussed in Sec.~\ref{Sec:single_to_single}. The critical line
converges to the $w=0$ point {\em linearly} ($\phi=1$) and  there
is no diverging time scale as $w$ becomes smaller with almost
perfect collapse of all critical density decay curves near
$w\approx 0$. Our crossover study provides another strong
numerical evidence that the {\tt tp12} belongs to the DP class.

\begin{figure}[b]
\includegraphics[width=0.45\textwidth]{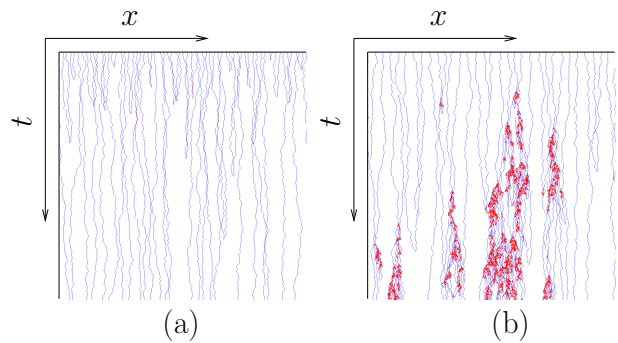}
\caption{\label{Fig:Spacetime} (Color online) Space-time
configurations of (a) the {\tt tp12} and (b) the PCPD at
criticality. The initial density is (a) $\frac{1}{16}$ for the
{\tt tp12} and (b) $\frac{1}{32}$ for the PCPD with size $L =
2^{10}$. Isolated particles (particles in a cluster) are
designated by the blue (red) color.}
\end{figure}
If the PCPD does not belong to the DP class, why
should the {\tt tp12} belong to the DP class? The PCPD  and the
{\tt tp12} seem quite similar in the sense of multi-particle
nature in reaction dynamics and also the absorbing space structure
(vacuum and a single-particle state). However, they are quite
different in the role of diffusing isolated particles.
See Fig.~\ref{Fig:Spacetime} for the space-time configurations for
the {\tt tp12} and the PCPD at criticality, starting from the low
density initial condition without pairs. The diffusing isolated
particles of the {\tt tp12} cannot increase the number of
particles in most cases, because $2 A \rightarrow \emptyset$
dynamics dominates over $3 A \rightarrow 4 A$ dynamics: Pairs
generated by collisions of two isolated particles evaporate before
greeting another isolated particle to become ``active'' triplets.
Consequently there is effectively no feedback mechanism from
isolated particles to increase the particle density or make the
system more active. Therefore the region of isolated particles can
be regarded as absorbing like in the PCP model. This case may
correspond to the {\em no-feedback} point ($r=0$) for the
generalized PCPD (GPCPD) studied in \cite{NP04}, which is the DP
fixed point.  This argument can be generalized to systems with
hybrid-type reaction dynamics $ m A \rightarrow (m+k)A$ and $ n A
\rightarrow (n-l) A$ with $m>n$ and $k,l>0$, which are numerically
shown to belong to the DP class \cite{KC03}.

The isolated particles of the PCPD, however, cannot be regarded as
absorbing as Fig.~\ref{Fig:Spacetime} shows; the isolated
particles may affect the critical spreading actively, because both
dynamics of $2 A \rightarrow \emptyset$ and $2 A \rightarrow 3A$
compete each other and consequently there is an effective feedback
mechanism from diffusing particles to make the system active. This
corresponds to the GPCPD with {\em long-term memory} effects at
$r\neq 0$ \cite{NP04}.

\section{\label{Sec:sum}Summary and conclusion}
In summary, we studied the crossover from the model belonging to
the directed percolation (DP) class with infinitely many absorbing
states (DP${}_\textrm{IM}$) to the DP model by reducing the number
of absorbing states significantly.  The crossover is found to be
well described by the usual crossover scaling function  for the
order parameter. The crossover exponent $\phi$ is argued to be
related to one of the DP critical exponent, {\it i.e.}, $\phi =
\nu_\|$, which is further evidenced by the similar crossover model
belonging to the directed Ising class. The origin of the diverging
scale and the nontrivial crossover comes from the discontinuity of
the auxiliary field density at the DP${}_\textrm{IM}$ critical
point. Our study for the first time presents the existence of the
nontrivial scaling in the DP${}_\textrm{IM}$, which is compared
with the study on the spreading exponents.

We also studied how the crossover behavior from the
DP${}_\textrm{IM}$ to the DP is affected by particle diffusion.
The crossover from the pair contact process with diffusion (PCPD)
to the DP studied in Ref.~\cite{PP06} is well classified by the
nontrivial crossover exponent. On the other hand, the {\tt tp12}
which is known to belong to the DP class is characterized by the
trivial ``crossover'' between the identical class detailed in
Sec.~\ref{Sec:single_to_single}. This provides an additional
evidence supporting that the PCPD is distinct from the DP.  In
addition, we suggest an argument based on the role of diffusing
isolated particles why the {\tt tp12} should belong to the DP
class, but the PCPD does not need to be.

It will be a challenging problem to see if the crossover scaling
from the DP${}_\textrm{IM}$ to the DP can be anticipated in the framework of
the field theory \cite{FT_PCP}.

\end{document}